\documentstyle[prb,aps,twocolumn,tighten,psfig,epsf,floats]{revtex}
\begin{document}
\draft
\title{{\em Ab initio} many-body calculations on infinite carbon and 
boron-nitrogen
chains}
\author{Ayjamal Abdurahman}
\address{Max-Planck-Institut f\"ur
Physik komplexer Systeme,     N\"othnitzer Stra{\ss}e 38
D-01187 Dresden, Germany}
\author{Alok Shukla} 
\address{ Physics Department, Indian Institute of
Technology, Bombay,   Powai, Mumbai 400076, India}
\author{Michael Dolg} 
\address{ Institut f\"ur Physikalische und Theoretische Chemie, Universit\"at Bonn,
Wegelerstr. 12, D-53115 Bonn, Germany}
 
\maketitle
\begin{abstract}
In this paper we report first-principles calculations on the ground-state
electronic structure of two infinite 
one-dimensional systems: (a) a chain of carbon atoms  and (b) a chain
of alternating boron and nitrogen atoms. 
Meanfield results were obtained using the restricted Hartree-Fock approach,
while the many-body effects were taken into account by 
second-order M{\o}ller-Plesset perturbation theory and the coupled-cluster 
approach. The calculations were performed using 6-31$G^{**}$ basis sets,
including the d-type polarization functions. 
Both at the Hartree-Fock (HF)
and the correlated levels we find that the infinite carbon chain exhibits 
bond alternation with alternating single and triple bonds, while the 
boron-nitrogen chain exhibits equidistant bonds. 
In addition, we also performed density-functional-theory-based local density approximation (LDA) calculations on the infinite carbon chain using the
same basis set. Our LDA results, in contradiction to our HF and correlated
results, predict a very small bond alternation. 
Based upon our LDA results for the 
carbon chain,  which are in agreement with an
earlier LDA calculation calculation [ E.J. Bylaska, J.H. Weare, and R. Kawai, Phys. Rev. B {\bf 58}, R7488 (1998).], we conclude that
the LDA significantly underestimates Peierls distortion. This emphasizes
that the inclusion of many-particle effects is very important for the
correct description of Peierls distortion in one-dimensional systems.
%In addition, 
%results for cohesive energies of these systems are also presented.

\end{abstract}

\section{Introduction}

Carbon occurs in several allotropic forms depending upon the
nature of its interatomic bonding. In diamond it exhibits $sp^3$ hybridization
resulting in a three-dimensional (3D) structure. In its graphite
form it exhibits $sp^2$ hybridization, resulting in a layered structure.  
While in the carbyne form, the carbon atoms exhibit $sp$
hybridization consistent with a predominantly one-dimensional (1D),
chain-like, character.~\cite{carbyne} Although, truly infinite isolated
single chains of carbon atoms do not exist in nature, however, there are
several experimental examples of the tendency
of carbon atoms to form 1D or quasi-1D structures. Lagow et al.~\cite{lagow} demonstrated that linear chains 
containing up to 28 atoms can be stabilized by adding nonreactive 
terminal groups. There is strong experimental evidence to suggest that
carbon chains and rings are also precursors to the formation of more
complicated structures such as fullerenes and 
nanotubes.~\cite{hunter,helden,kiang} In addition, atom resolved scanning
tunneling microscopy experiments on
the (100) surface of $\beta$-SiC indicate the formation of carbon
atomic chains.~\cite{derycke} 
Similarly, simulations performed on tetrahedral amorphous
carbon surfaces also indicate the formation of rings and chains.~\cite{dong}
Additionally, for the carbon cluster anions C$_{n}^{-}$ there is 
sufficient experimental evidence to indicate the existence of chains for
$n \le 10$, and rings for the larger structures.~\cite{anions}

   From a theoretical point of view as well carbon chains and rings have
elicited a large amount of 
interest.~\cite{mitas,saito,bylaska,fuent,jones,lou,martin,eriksson,watts,%
malek,springborg,rice,teramae,karpfen,kertesz1,kertesz2} Let us assume an infinitely long 
carbon chain (C chain hereafter), with the conjugation direction along the $x$-axis. If the
chain were to have equal bond distance (cumulenic character with exclusively
double bonds), then it will exhibit metallic character with the
occupied band structure consisting  of: (a) a core band
composed of 1s orbitals (b) a valence $\sigma$ band consisting mainly of
sp-hybrids and (c) two half-filled degenerate $\pi$ bands at the Fermi level
composed of $\pi_y$ and $\pi_z$ orbitals.~\cite{eriksson,springborg,karpfen,%
kertesz1} 
Therefore, the question arises whether such a chain will undergo a Peierls-type
distortion to become an insulator with bond alternation along the chain
(acetylenic character with alternating single and triple bonds). For a
finite ring of carbon atoms the same question reads: will it exhibit
aromatic (equal bonds) or antiaromatic (bond alternated) 
behavior?~\cite{mitas,saito,bylaska,martin} Carbon rings containing
$4N$ atoms ($N$ is an integer) undergo bond alternation due to first-order
Jahn-Teller (JT) distortion.~\cite{mitas,saito,bylaska,martin,salem} However,
the answer for a ring containing $4N+2$ atoms is not straightforward
because, if it undergoes bond alternation, it can only be due to a second
order JT effect.~\cite{mitas,saito,bylaska,rem1} Perhaps it is because of
this uncertainty as to whether the infinite linear C chain 
(C$_{4N+2}$ ring, with large $N$) will undergo Peierls distortion (second-order
JT effect) which has led 
to numerous theoretical studies of these systems.~\cite{mitas,saito,%
bylaska,fuent,jones,lou,martin,eriksson,watts,malek,springborg,rice,teramae,%
karpfen,kertesz1,kertesz2,theory}
First principles calculations of the infinite C chain fall in two
categories, namely, the Hartree-Fock 
(HF)~\cite{teramae,karpfen,kertesz1,kertesz2}, 
and the density functional theory (DFT)
calculations.~\cite{saito,bylaska,eriksson,malek,springborg} 
{\em Ab initio} HF calculations invariably predict an infinite C chain
with bond alternation to be energetically more stable as compared to
the equidistant configuration.~\cite{teramae,karpfen,kertesz1,kertesz2}
However, HF calculations generally have a tendency to overestimate bond
alternation which was also evident in our previous studies on 
{\em trans}-polyacetylene.~\cite{yuming,tpa-shukla} It is only the subsequent treatment
of electron correlation effects which brings the bond-alternation in such
systems closer to the experimental reality.~\cite{yuming}
Therefore, one would expect that the DFT-based approaches, which do include
electron-correlation effects in an approximate manner, would provide an
unambiguous picture. However, this is not the case. Whereas the local density
approximation (LDA) results of Springborg and
collaborators~\cite{eriksson,malek,springborg} place the 
bond alternation in the range  
$0.1 \AA \leq \delta \leq 0.2 \AA$ ($\delta = r_{single}-r_{triple}$), the recent
LDA study of Bylaska et al.~\cite{bylaska} predicts only a very small
value $\delta < 0.04 \AA$ for the infinite chain. Keeping in view these
conflicting results obtained for the bond alternation from the DFT
calculations, and the fact that HF results tend to overestimate $\delta$,
we believe that a systematic study of the influence of electron correlation
effects  on the bond alternation in the infinite C chain is in order.
Therefore, we decided to apply our {\em ab initio}
wave-function-based many-body methodology used earlier to study a
number of conjugated polymers,~\cite{yuming,ayj-bn,ayj-lih,ayj-pmi}
to the problem of the infinite C chain as well. In addition, to put our
correlated results in proper perspective, we also performed LDA calculations
on the C chain using the same basis set which was used for the HF and
the subsequent many-body calculations. These LDA calculations were 
performed using the CRYSTAL98 program~\cite{crystal1}.
Indeed our many-body calculations predict
$\delta  \approx 0.15 \AA$ close to the findings 
of Springborg et al.,~\cite{eriksson,malek,springborg} 
but in complete disagreement with those of Bylaska et al.~\cite{bylaska}
However, our LDA calculations performed with the CRYSTAL98
code~\cite{crystal1} predict a very small bond alternation of
$\delta  \approx 0.04 \AA$, in excellent agreement with the LDA
value of $\delta  \approx 0.036 \AA$ reported by Bylaska et al.~\cite{bylaska}

Another useful concept to quantify the tendency towards
bond alternation is the condensation energy defined as the  difference in 
energy/atom of the equidistant configuration and the bond alternating
configuration. Our best value of the condensation energy, obtained at the
many-body  level using the coupled-cluster approach, is 18.4
milliHartrees/atom. This value is much larger than 
than 0.13 milliHartrees/atom predicted by our LDA calculations and
also the LDA value of 0.09 milliHartrees/atom, reported by 
Bylaska et al.~\cite{bylaska}
Thus our many-body calculations indicate a much stronger tendency for the
infinite carbon chain to undergo Peierls distortion as compared to the LDA 
calculations.

    In addition to the infinite C chain, in this paper we also study
an infinite chain composed of boron and nitrogen atoms (BN chain, hereafter).
If we consider a two atom unit cell, the carbon  and the BN chains are
isoelectronic. Thus if the two chains exhibit different behavior, it will be 
entirely due to the differences in the nature of bonding caused by different
atomic species. As a matter of fact, starting from well-known carbon-based
materials, the prospect of obtaining novel material by performing
alternate boron and nitrogen substitution, has fascinated experimentalists 
and theoreticians alike.~\cite{ayj-bn,armstrong,paine,rubio,blase,chopra,%
march,cote} Of course, cubic and hexagonal BN, which are the
counterparts of diamond and graphite, respectively, are well-known 
materials.
In our own earlier work,~\cite{ayj-bn} as also in the works of Armstrong and 
collaborators,~\cite{armstrong} the BN analogs of {\em trans}-polyacetylene
and polyethylene were investigated theoretically. BN analogs of fullerene-like
cages and nanotubes have been  studied theoretically by several 
authors.~\cite{rubio,blase,march}
Experimentally, the BN nanotubes were first synthesized by Chopra 
et al.,~\cite{chopra} while recently, multiwall BN nanotubes, as well as
fullerene-like BN cages, have been synthesized by Lee et al.,~\cite{lee} in a 
catalyst-free manner, using the laser-ablation technique.
Additionally, C\^{o}t\'{e} et al.~\cite{cote} calculated the
optical properties of the BN analogs of photoluminescent conjugated polymers
poly(para-phenylene) (PPP) and poly(para-phenylenevinylene) (PPV). However, 
to the best of our knowledge, the infinite isolated BN chain has not been 
studied theoretically so far.
Therefore, in light of the current interest in novel BN compounds structurally
similar to carbon-based compounds, a parallel study of the infinite BN chain
is desirable. Indeed, we find that the infinite BN chain has properties
different than the infinite C chain in that it does not exhibit
bond alternation. This result is similar to our earlier result where
we found that the BN analog of {\em trans}-polyacetylene does not exhibit
bond alternation.~\cite{ayj-bn}

The remainder of this paper is organized as follows. In section \ref{method} 
the applied methods and computational details are briefly described. 
The results are 
then presented and discussed in section \ref{results}. Finally, our
conclusions are presented in  
section \ref{conclusions}.     
\section{Applied Methods and Computational Details.}
\label{method}
For the linear infinite C chain we chose a unit cell
consisting of two carbon atoms, while for the BN chain a corresponding unit 
cell
consisting of one boron atom and one nitrogen atom was considered.
For both systems HF calculations were performed first to
obtain the results at the mean-field level. For the dimerized C chain,
as well as for the BN chain, the HF calculations were performed in the
infinite system limit using both the Bloch-orbital-based electronic
structure program CRYSTAL,~\cite{crystal1,crystal2} as well as our own
Wannier-function-based program WANNIER.~\cite{tpa-shukla} 
Both these programs use the linear-combination of atomic orbital (LCAO) 
approach based on Gaussian-type basis functions. However, for the
equidistant C chain, we encountered convergence difficulties with
the CRYSTAL program when diffuse basis functions were used. Therefore,
for the metallic C chain, we obtained HF results by performing finite
cluster calculations using clusters of increasing size, and by ensuring that
the convergence with respect to the cluster size was achieved.
 To further
check the accuracy of the finite-cluster approach, the  
finite-cluster HF calculations were also repeated for the case of insulating 
chains for which the infinite system HF results (CRYSTAL and WANNIER) were
available, and excellent agreement between the two sets of results was
obtained. 

As mentioned earlier, for the purpose of comparison with other approaches,
 we also performed 
LDA calculations on the infinite C chain to study its bond-alternation
properties. These LDA calculations were performed with the 
CRYSTAL98 program~\cite{crystal1},
employing the same  6-31$G^{**}$ basis set as was used in the HF and
 the subsequent many-body calculations. 
Both in the HF and the LDA calculations performed with the Bloch-orbital-based
CRYSTAL program,~\cite{crystal1,crystal2} particular attention was paid 
to the convergence of total energies with respect to the number of
 ${\bf k}$-points used in the Brillouin-zone (BZ) integration. The number of
${\bf k}$-points was increased until the total energy/cell had 
converged at least up to $1.0 \times 10^{-6}$ Hartrees. We
 noticed that using thirty ${\bf k}$-points in the
 irreducible part of the BZ was sufficient to achieve aforesaid convergence
both for HF and the LDA calculations performed with the 
CRYSTAL program.~\cite{crystal1,crystal2} Moreover, the excellent agreement
observed among the HF energies/cell computed using the CRYSTAL program,
our own WANNIER program,~\cite{tpa-shukla} and the finite-cluster-based HF calculations, leaves
no doubt that the convergence in energies has been achieved.

The many-body calculations beyond HF were 
performed only within the finite-cluster model. However, since the correlation
effects are highly localized in real space, convergence with respect to the
cluster size is rarely a problem here. Moreover, in our earlier calculations
on the infinite LiH chain,~\cite{ayj-lih} and the bulk LiH,~\cite{lih-shukla}
we had carefully compared the correlated calculations performed simultaneously
on the infinite system and its finite cluster, and found excellent agreement
between the two sets of results. The issue of the convergence of the present
set of results with respect to the cluster size is very important, and, therefore, will be discussed again in
section \ref{results}. 

In order to perform the correlated calculations,
three different many-body approaches, viz., second-order M{\o}ller-Plesset
perturbation theory (MP2), coupled-cluster singles and doubles (CCSD), and
coupled-cluster singles and doubles with perturbative treatment of the
triples (CCSD(T)) were used. All the finite-cluster-based calculations,
both at the HF and the correlated level, were performed with the 
MOLPRO  molecular orbital {\em ab initio\/} program 
package.~\cite{molpro} In order to minimize the boundary effects associated
with the finite clusters, we extracted the energy per unit cell from the
finite-cluster calculations by computing the total energy differences 
between the clusters containing $n$, and $n+1$ unit cells, with as large
a value of $n$ as feasible. In view of the typical accuracy of the periodic
HF codes, i.e., 1 milliHartree per atom, we consider the results to be converged 
with respect to $n$, when the (total/correlation) energy per unit cell is stable 
within this accuracy. The undimerized C chain, which consists of C atoms
connected by double bonds, was terminated with two H atoms
on each end, and the energy/cell was computed using the expression
\begin{equation} 
E=\lim_{n \to \infty}\bigtriangleup E_{n}=\lim_{n \to
\infty}\biggl[E(C_{2n+2}H_{4})-E(C_{2n}H_{4})\biggr] .
\label{eq-cudim}
\end{equation} 
The insulating C chain consisting of alternating single and triplet bonds,
 was terminated with a single CH bond, 
leading to the formula
\begin{equation} 
E=\lim_{n \to \infty}\bigtriangleup E_{n}=\lim_{n \to
\infty}\biggl[E(C_{2n+2}H_{2})-E(C_{2n}H_{2})\biggr].
\label{eq-cdim}
\end{equation}
As far as  the BN chain is concerned, its bonding pattern is not obvious.
Therefore, several end-termination schemes to saturate the dangling bonds
were explored. These included terminating the chains with: (a) one
H atom on each
end (b) two H atoms on each end, and (c) unequal number of H atoms on the two
ends. However, converged energies per cell were found to be quite insensitive
to the end geometry of the chain. Therefore, in close analogy with the
undimerized C chain,  
we finally settled with terminating  the BN chain with two H atoms on 
each end,  and the  expression
\begin{equation}
E=\lim_{n \to \infty}\bigtriangleup E_{n}=\lim_{n \to
\infty}\biggl[E((BN)_{n+1}H_{4})-E((BN)_{n}H_{4})\biggr]
\end{equation}
was used to compute its energy per unit cell. As discussed in the
next section,  the HF energy/cell
of the BN chain obtained by this method is in excellent agreement with
the ones obtained from the infinite-chain calculations performed with
the CRYSTAL and WANNIER programs. Therefore, we believe that the 
aforesaid termination scheme, which effectively models a BN chain as
composed of double bonds, is a sound one.

\section{Results and Discussions}
\label{results}
Calculations on both the systems were performed with 6-31$G^{**}$ basis sets. 
For the HF calculations on the infinite insulating chains, we had to increase
the outermost p-type exponents of the original basis sets because of the
linear-dependence related problems.
This led to
$0.29871$ for carbon (original 0.16871), $0.16675$ for 
boron (original 0.12675), and $0.29203$ for nitrogen (original 0.21203).
The outermost s-type exponents of the original basis set were left unaltered.
The d-type exponents used for each atom were:
0.8 for carbon, 0.6 for boron, and 0.8 also for nitrogen. These d exponents
were preferred over the original d-type exponents available from the
MOLPRO library,~\cite{molpro} because they led to lower total energies. 
For the correlated
calculations, we used the original basis set to compute the energy/cell 
and the correlation energy was computed by subtracting from it the HF 
energy/cell obtained by finite-cluster calculations done also with the
original basis set. In this manner, we ensured that no spurious contributions
to the correlation energy were obtained because of the use of two different
basis sets during these calculations. Next we discuss the results obtained for
the two systems in detail.

\subsection{Carbon Chain}
\label{c-chain}
Our final results for the equidistant and dimerized C chains, employing the
wave-function-based approaches (HF, MP2, CCSD, CCSD(T)), are
presented in tables \ref{c-udim} and \ref{c-dim}, respectively. Results
of our LDA calculations performed with CRYSTAL98 program,~\cite{crystal1}
along with the comparison of our results with those of other authors, 
are presented in table \ref{tab-ccomp}. As mentioned
earlier we experienced convergence difficulties with the CRYSTAL program
for the undimerized chain, so HF results for that system were obtained
by the finite-cluster approach only. Moreover, all the correlated calculations
were also done in a finite-cluster model. Therefore, it is important for us to
demonstrate the convergence of all the finite-cluster-based results  
for both the undimerized, and the dimerized C chain, as
a function of the cluster size. 

\begin{figure}
\caption{Total HF energy per unit cell for both the undimerized (circles) and 
dimerized (squares) carbon 
chains computed by the finite-cluster approach 
(cf. Eqs. (\protect\ref{eq-cudim}), and (\protect\ref{eq-cdim})), plotted as a function
of the number of unit cells $n$. Two carbon atoms were assumed per unit cell.
The bond distances in these calculations correspond to the optimized
geometries reported in tables \protect\ref{c-udim} and \protect\ref{c-dim}. }
\psfig{file=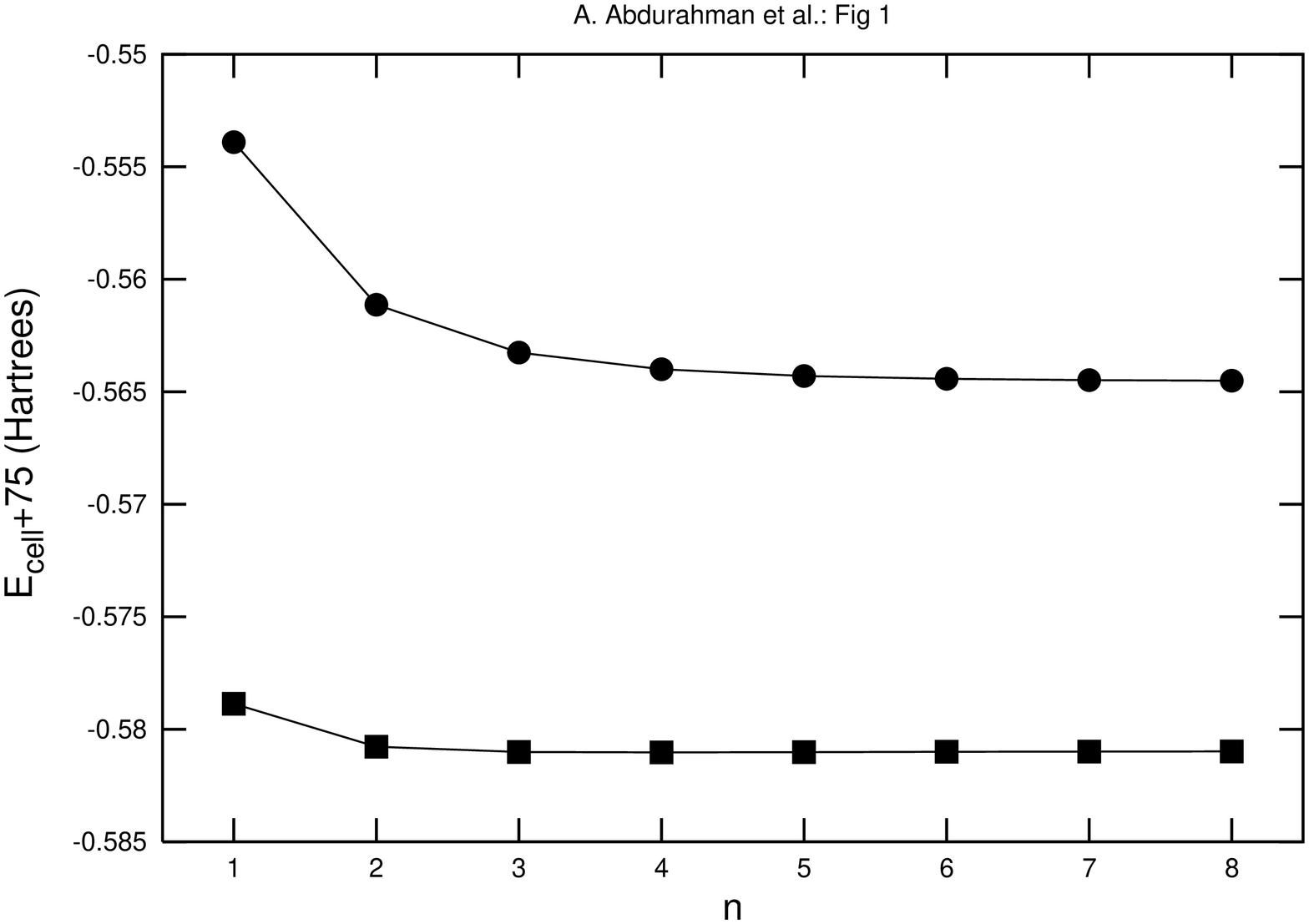,width=10.cm,angle=-90}
\label{fig-c-scf}
\end{figure}

\begin{figure}
\caption{Correlation energy per unit cell computed by the CCSD(T) approach
both for undimerized (circles) and dimerized (squares) finite carbon chains, plotted as a function of the 
number of unit cells $n$ in the finite cluster. For the undimerized chain $r_{double} = 1.26 \AA$, 
and for the dimerized chain $r_{single} = 1.37 \AA$, and $r_{triple}=1.175 \AA$.} 
\psfig{file=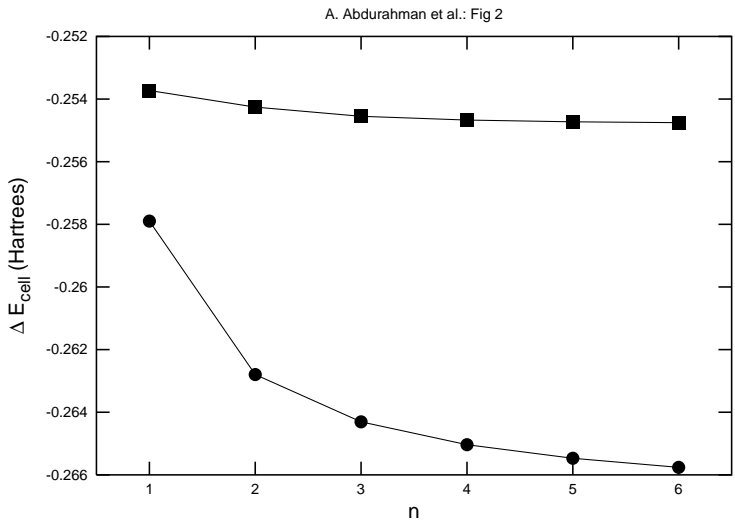,width=10.0cm,angle=-90}
\label{fig-c-cor}
\end{figure}

Total HF energy/cell as a function of the 
number
of unit cells $n$ in the cluster is plotted in Fig. \ref{fig-c-scf}, and
it is clear that the convergence with respect to the cluster size has been 
achieved both for the dimerized and undimerized chains, at least, to the milliHartree level. Table \ref{c-udim} also presents the total
energy/cell for the undimerized chain including contributions of the electron
correlation effects, while the corresponding data for the dimerized chain
is presented in table \ref{c-dim}. 
The convergence of the correlation energy/cell computed by the CCSD(T) method
for both types of chains, with respect to the cluster size, is presented in
Fig. \ref{fig-c-cor}. It is clear from these figures that the energies for the
dimerized chain converge very rapidly, consistent with our intuitive picture
of localized electrons in such systems. The convergence in the undimerized
chain, although comparatively slower, has also
been achieved, at least, to the milliHartree level. 

The optimized geometries
reported in this work were obtained by the usual
procedure of first performing several total energy calculations for various 
geometry parameters, and then fitting the results by a least-squares procedure
to polynomials of suitable degrees. The fact that the finite-cluster results
at the HF level agree fully with those obtained by the CRYSTAL and WANNIER
programs gives us added confidence in their correctness.
It is clear that the dimerized structure is energetically much more stable
as compared to the metallic one, at all levels of calculations (SCF, MP2, CCSD,
CCSD(T)). At the HF level we obtained a percentage bond alternation
 $\delta(\%)$=14.5\% (where $\delta(\%) = (\delta/r_{av}) 
\times 100$, with $r_{av} = (r_{single} + r_{triple})/2$), and at the CCSD(T)
level of correlation it reduces to
11.8\%. The condensation energy, defined earlier as the difference
in the total energy/atom between the optimized undimerized and the dimerized
C chains, at various levels of correlation treatments, is obtained to be 
$E_{cond}$(HF) = 8.8 milliHartrees, $E_{cond}$(MP2) = 19.25 milliHartrees,
$E_{cond}$(CCSD) = 20.35 milliHartrees, and $E_{cond}$(CCSD(T)) = 18.4
milliHartrees. Thus, based on the CCSD(T) approach, which is the most sophisticated correlation approach
that we have used in these calculations, our final result for the
condensation energy of an infinite C chain of 18.4 milliHartrees/atom.
As reported in table \ref{tab-ccomp}, the value of condensation energy
obtained from our LDA calculations performed with CRYSTAL98
program~\cite{crystal1} was 0.13 milliHartrees/atom, which is very close to  
the LDA value of 0.09 milliHartrees/atom reported by
Bylaska et al.,~\cite{bylaska} and much smaller than our many-body results
based upon the CCSD(T) method. 

The summary  of our many-body and LDA results, along  with those
obtained by other authors is presented in Table \ref{tab-ccomp}.
It is clear from the table that the results obtained by various authors on the
optimized
geometry parameters of the C chain vary quite a bit, however, almost all
of them report the bond alternation to be more than 10\%. The exceptions
to this result are (a) our own LDA calculation ($\delta(\%) = 1.6$), (b) 
LDA calculations of Eriksson et al.~\cite{eriksson} ($\delta(\%) = 6.0$), and
(c) LDA calculations of Bylaska et al.~\cite{bylaska}($\delta(\%) = 2.9$).
Thus these recent LDA results, which predict rather small bond alternation
for the C chain, are in contradiction with the older LDA 
calculations~\cite{malek,springborg}, as well with our many-body  results.
We believe that one should expect a rather small bond alternation from the
LDA based calculations, because the homogeneous-electron-gas approximation
therein leads preferentially to a metallic ground state.
At this point, we would like to mention that recently Torelli 
et al.~\cite{mitas} performed calculations on the optimized ground state
geometries of C$_{4N+2}$ rings using a quantum Monte Carlo method and their 
results for the bond alternation for the C$_{18}$ and C$_{22}$ rings 
were 7\%. Although,
our final CCSD(T) result of 11.8\% bond alternation for the infinite C chain
is more than that obtained by Torelli et al.~\cite{mitas} for the large
C rings, yet we believe that qualitative trends indicated by most of the
calculations suggest a significant bond alternation in the infinite C chains.
In this context we would like to mention that for the finite 
ring systems containing $4N+2$ carbon atoms, the H\"uckel theory predicts that
the bond alternation due to the second-order Jahn-Teller distortion
will increase with increasing $N$, eventually coalescing with the value
corresponding to  the Peierls distortion, for 
$N \rightarrow \infty$.~\cite{klein} Discounting the basis-set effects,
in our opinion, our value of bond alternation of 11.8\% for the  infinite
C chain, as compared to 7\% reported by Torelli et al.~\cite{mitas} for
the C$_{22}$ rings, is indicative that the aforesaid H\"uckel result is
valid even when the many-body effects have been included.

 Other authors have not reported the value of the condensation energies
obtained in their calculations. But, based on the trends visible from the
calculations on bond alternation, we believe that it is safe to assume that
the condensation energy for the infinite C chain is in access of 10
milliHartrees/atom. Thus for condensation energy, our many-body results, 
 as well that of the Torelli et al.~\cite{mitas} for the finite rings,
are quite different from our LDA results of 0.14 milliHartree/atom, and 
0.1 milliHartree/atom reported by Bylaska et al.~\cite{bylaska}

\subsection{BN Chain}
\label{bn-chain}
\begin{figure}
\caption{Correlation energy per unit cell of finite boron-nitrogen chains computed by the
CCSD(T) approach, plotted as a function of the number of unit cells
$n$ in the finite cluster. The bond distance was taken to be $r_{BN} =1.301 \AA$.} 
\psfig{file=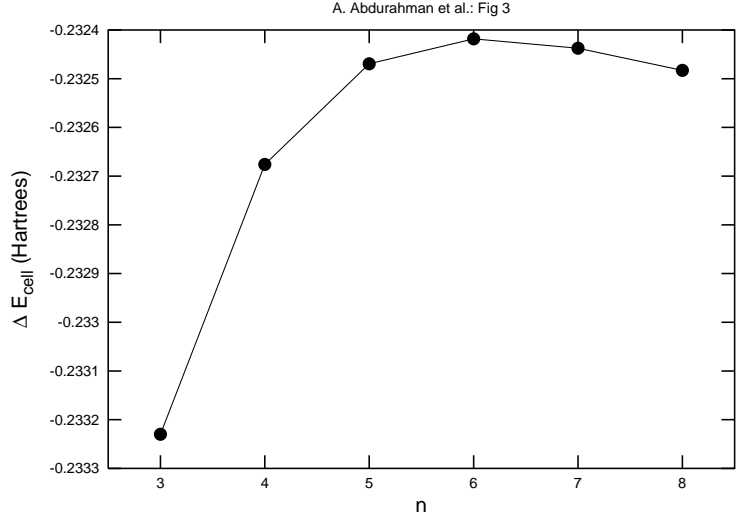,width=10.0cm,angle=-90}
\label{fig-bn-cor}
\end{figure}

As mentioned earlier, in nature there appears to be a one-to-one correspondence
between several structures of carbon and BN allotropes such as diamond vs. 
cubic BN, graphite vs. hexagonal BN, and carbon nanotubes vs. BN nanotubes etc.
Therefore, it will be natural to assume that a BN chain, similar to a carbon 
chain, will also have a strictly linear geometry exhibiting sp-type
hybridization with, 180$^o$ bond angles.  
However, this being the first theoretical study of this system, we decided
to be careful with this assumption. Therefore, in our calculations, along 
with the linear geometry, we also  explored the energetics of various 
zig-zag structures of the BN chain. However, we found that in all cases
the zig-zag structure led to higher energies compared to the linear one,
leading us to the conclusion that if the BN chains exist, similar to 
the C chains, they will have a linear structure. Table \ref{tab-bn} presents
the results of our calculations for the BN chain both at the HF and the
correlated levels. Since we obtained perfect agreement on the HF energy/cell
for the BN chain from WANNIER, CRYSTAL, and finite-cluster-based approaches,
we are again very confident of the correctness of our correlated studies
performed on finite BN clusters saturated with hydrogens on the ends. 
To further confirm this point, in Fig. \ref{fig-bn-cor} we present the
variation of the correlation 
energy/cell of BN chain, as a function of the number of unit cells 
in the cluster. It is obvious from the figure that satisfactory convergence
in correlation energy has been achieved. At the CCSD(T) level we obtain
a cohesive energy of 11.72 eV/cell, which is somewhat greater than the
value of 11.05 eV/cell obtained for the dimerized C chain. This means
that from an energetic point of view, the BN chain should at least be as stable
as an isolated C chain, and, therefore, the probability of
finding linear sp-type hybridized structures composed of BN units should be 
as high as that of the C units. 

The other remarkable aspect of the
BN chain, as compared to the C chain, is that it does not exhibit any
bond alternation. This result is identical to our earlier theoretical study
on polyiminoborane, which is the BN substituted analog of 
{\em trans}-polyacetylene.~\cite{ayj-bn} In contrast to {\em trans}-polyacetylene which
exhibits a bond alternation close to 0.1 $\AA$, in case of polyiminoborane we
 obtained optimized geometries with equidistant 
BN atoms.~\cite{ayj-bn} 
Of course, the Peierls theorem~\cite{peierls} does not apply to BN chain or 
polyiminoborane, because these systems are insulators even without bond
alternation.~\cite{insul} 
However, the absence of bond alternation in 
BN compounds can be understood from a different line of reasoning as well. 
BN systems, being heteronuclear in nature, 
have significant ionic contributions to bonding as opposed to the 
purely covalent carbon-based materials. Therefore, in
BN systems in addition to the covalent interactions, 
electrostatic attraction between the B and N sites will also exist because
of a degree of ionicity. It is this electrostatic interaction due to
ionicity in the BN systems which, in our opinion, leads to equal bond lengths. 
The optimized BN distance at the CCSD(T) level was obtained to be 1.3 $\AA$,
which is slightly more than the optimized average bond distance of 1.28 $\AA$
obtained for the infinite C chain. For the linear chain, our optimized 
BN distance of 1.3 $\AA$ should be compared to the in plane BN distance of 
1.45 $\AA$ in the hexagonal BN, and the BN distance of 1.57 $\AA$ observed in 
the cubic BN. The significant difference in these three BN distances is
consistent with the fact that the BN chain, hexagonal BN and the cubic BN
originate from very different bonding schemes based on sp, sp$^2$, and
sp$3$ hybridizations, respectively.

\section{Conclusions}
In conclusion, we have presented an {\em ab initio} study of the ground state
properties of the infinite carbon and boron-nitrogen chains. The influence of
electron correlation effects was included by several approaches including
the coupled-cluster method accounting for up to triple excitation operators. 
As far as the carbon chain is concerned, all our many-body calculations
predict a dimerized ground state, consistent with the phenomenon of Peierls
distortion. Thus our calculations on the infinite C chain based upon many-body methods such as the
coupled-cluster approach are in agreement with the Monte Carlo calculations
of Torelli et al.~\cite{mitas} performed on carbon rings, which also revealed
a strong tendency of these systems to exhibit bond alternation in the
infinite chain limit. We also performed a parallel LDA calculation on the
infinite C chain which predicted very weak dimerization, in excellent 
quantitative agreement with the recent LDA results of Bylaska et 
al.~\cite{bylaska} This, in our opinion, establishes in most unambiguous
manner that LDA underestimates Peierls distortion in one-dimensional
systems, and that powerful many-body methods such as the couple-cluster
approach are essential for the correct description of the broken-symmetry
ground state.

 In addition, we also presented first principles correlated calculations on
the ground state of an infinite boron-nitrogen chain. Based upon our calculations we
conclude that the boron-nitrogen chain has a higher cohesive energy/cell as compared to the
carbon chain, thus making BN-based 1D-structures energetically at least as 
favorable as the carbon-based ones. Of course, it is a difficult task to synthesize
isolated infinite atomic chains. However, similar to the case of carbon, it
should certainly be possible to test in the laboratory whether finite 
one-dimensional clusters based on BN units exist. Along the same lines,
it will also be worthwhile to theoretically optimize the geometries
of the ground states of various BN-based finite clusters. In addition to the
ground state, theoretical studies of their optically excited states will also
be useful because they, in conjunction with photoabsorption-based experiments,
can be used to characterize various clusters. We will investigate the
electronic structure of BN clusters in a future publication.

\label{conclusions}
\acknowledgements
A.A. would like to express her gratitude to Prof. P. Fulde for supporting her
research work. Part of this work was performed when A.S. was a 
guest scientist in the Quantum Chemistry group of the
Max-Planck-Institut f\"ur Physik Komplexer Systeme, Dresden, Germany. 
He is grateful to the group for their hospitality. Authors would also like
to thank Drs. U. Birkenheuer and S. Pleutin for reading the manuscript
critically, and for suggesting improvements.

\clearpage
\newpage
%\clearpage
%\newpage
%
\clearpage
%\newpage
%
%\clearpage
%\newpage
%
\begin{table}
\caption{Total energy E$_{tot}$ (Hartree), cohesive energy $\bigtriangleup$E$_{coh}$ (eV) 
 per unit $(C-C)$ and bond lengths$(\AA)$ for the undimerized structure of 
the infinite carbon chain. Cohesive energies were computed with respect to
separated atoms.}
\begin{tabular}{lllll}
\hline
Method & E$_{tot}$ & $\bigtriangleup$E$_{coh}$&$r_{double}$  \\
\hline
\hline
%CRYSTAL SCF& -75.5574 & 4.893 & 1.253     \\
Finite cluster SCF&-75.5644&5.733&1.251\\
MP2&-75.7847&10.327&1.268 \\
CCSD & -75.7813&9.570&1.267 \\   
CCSD(T)&-75.7996&10.052&1.272 \\
\hline
\hline
\end{tabular}
\label{c-udim}
\end{table}
\clearpage
\newpage
\begin{table}
\caption{Total energy E$_{tot}$ (Hartree), cohesive energy $\bigtriangleup$E$_{coh}$ (eV) 
 per unit $(C-C)$ and bond lengths$(\AA)$ for the dimerized structure of the infinite
carbon chain. Cohesive energies were computed with respect to
separated atoms. }
\begin{tabular}{lllll}
\hline
Method & E$_{tot}$ & $\bigtriangleup$E$_{coh}$&$r_{single}$&$r_{triple}$  \\
\hline
\hline
WANNIER SCF&-75.5807&6.176&1.367&1.173     \\
CRYSTAL SCF&-75.5798&6.150&1.369&1.173     \\
Finite cluster SCF&-75.5810&6.184&1.360&1.174   \\
MP2& -75.8232&11.375&1.337&1.217\\
CCSD & -75.8220&10.678&1.362&1.197\\
CCSD(T)&-75.8364&11.053 &1.358&1.207\\
\hline
\hline
\end{tabular}
Correlation contributions added to CRYSTAL SCF energies.\\
\label{c-dim}
\end{table}
%\clearpage
%\newpage
%
\begin{table}
\caption{Infinite carbon chain: comparison of the present results with those of other 
authors. The percentage bond alternation $\delta(\%) = (\delta/r_{av}) \times 100$, where
$\delta = r_{single} - r_{triple}$, and $r_{av} = (r_{single} + r_{triple})/2$.}
\begin{tabular}{llllll}
\hline
Author & Method   & $r_{single}$ & $r_{triple}$ & $\delta(\%)$    & $E_{cond}$ \\
       &          & ($\AA$)      & ($\AA$)      &    & (mHartrees/atom) \\
\hline
\hline
This work             &  RHF     & 1.360     & 1.174     & 14.5 & 7.80   \\ 
This work             &  CCSD(T) & 1.358     & 1.207     & 11.8 & 18.40  \\
This work             &  LDA     & 1.286     & 1.246     & 1.6  & 0.13  \\
Bylaska et al.~\cite{bylaska}     & LDA      & 1.288$^a$ & 1.252$^a$ & 2.9 $^b$ & 0.09$^b$ \\
Eriksson et al.~\cite{eriksson}   & LDA      & 1.376     & 1.296     & 6.0 & --- \\
Springborg et al.~\cite{malek} & LDA      & 1.503     & 1.259     &    18.0 & --- \\ 
Springborg~\cite{springborg}        & LDA      & 1.439     & 1.249     &  14.1   & --- \\
Karpfen~\cite{karpfen}           & RHF      & 1.363     & 1.198     &   12.9  &  --- \\
Kert\'{e}sz et al.~\cite{kertesz1} & RHF     & 1.145$^c$ & 1.405$^c$ & 20.4   &  --- \\ 
Kert\'{e}sz et al.~\cite{kertesz2} & UHF     & 1.335     & 1.185     & 11.9  & --- \\
Teramae et al~\cite{teramae}.     & RHF     & 1.339     & 1.166     & 14.1 &  --- \\ 
\hline
\hline
\end{tabular}
\label{tab-ccomp}
$^a$ Calculated from the average bond distance of 1.270 \AA \ and $\delta$ = 2.9 \%. \\
$^b$ Upper limit (Ref.~\cite{bylaska}) \\
$^c$ Calculated from the average bond distance of 1.275 \AA \ and $\delta$ = 20.4 \%. \\
\end{table}
\clearpage
%\newpage
%
\begin{table}
\caption{Total energy E$_{tot}$ (Hartree), cohesive energy
$\bigtriangleup$E$_{coh}$ (eV) per unit $(BN)$ and 
 B--N bond length r$_{BN}$ $(\AA)$ of the infinite boron-nitrogen chain.
Cohesive energies were computed with respect to
separated atoms. }
\begin{tabular}{llll}
\hline
Method & E$_{tot}$ & $\bigtriangleup$E$_{coh}$&r$_{BN}$  \\
\hline
\hline
WANNIER SCF&-79.1801&7.5845&1.287\\
CRYSTAL SCF &-79.1794&7.5656&1.288\\
Finite cluster SCF&-79.1773&7.5076&1.287  \\
MP2& -79.3968&11.9478&1.300\\
CCSD& -79.4017&11.4723&1.299\\
CCSD(T)&-79.4113&11.7204&1.301\\
\hline
\hline
\end{tabular}
Correlation contributions added to CRYSTAL SCF energies.\\
\label{tab-bn}
\end{table}


\begin{references}
\bibitem{rem1} Clearly, for a very large ring 
($N \rightarrow \infty$) the second order JT effect of the $4N+2$ ring will
become identical to the Peierls distortion of the infinite chain.
\end{references}
\end{document}